\documentclass[sigconf]{acmart}
\usepackage{booktabs} 
\usepackage{amsmath,graphicx}
\usepackage{tabularx}
\usepackage{todonotes}
\usepackage{float}
\def\x{{\mathbf x}}

\setcopyright{rightsretained}
\newcommand\cev[1]{\overleftarrow{#1}}

\acmDOI{10.475/123_4}

\acmISBN{123-4567-24-567/08/06}

\acmConference[WOODSTOCK'97]{}{}{} 
\acmYear{1997}
\copyrightyear{2016}

\acmPrice{15.00}

\begin{document}
\title{Semi-Supervised Recurrent Neural Network for Adverse Drug Reaction Mention Extraction}

\author{Shashank Gupta}
\authornote{Work done during internship at TRDDC, Pune.}

\affiliation{%
  \institution{International Institute of Information Technology}
  \streetaddress{Hyderabad, India}
}\email{shashank.gupta@research.iiit.ac.in}

\author{Sachin Pawar}
\affiliation{%
  \institution{TCS Research, Tata Consultancy Services}
  \streetaddress{Pune, India}
}
\email{sachin7.p@tcs.com}

\author{Nitin Ramrakhiyani}
\affiliation{%
  \institution{TCS Research, Tata Consultancy Services}
 \streetaddress{Pune, India}
}
\email{nitin.ramrakhiyani@tcs.com}

\author{Girish Keshav Palshikar}
\affiliation{%
  \institution{TCS Research, Tata Consultancy Services}
 \streetaddress{Pune, India}
}
\email{gk.palshikhar@tcs.com}

\author{Vasudeva Varma} 
\affiliation{%
  \institution{International Institute of Information Technology}
  \streetaddress{Hyderabad, India}
}\email{vv@iiit.ac.in}

\renewcommand{\shortauthors}{B. Trovato et al.}

\begin{abstract}
Social media is an useful platform to share health-related information due to its vast reach. This makes it a good candidate for public-health monitoring tasks, specifically for pharmacovigilance. We study the problem of extraction of Adverse-Drug-Reaction (ADR) mentions from social media, particularly from Twitter. Medical information extraction from social media is challenging, mainly due to short and highly informal nature of text, as compared to more technical and formal medical reports. 

Current methods in ADR mention extraction rely on supervised learning methods, which suffer from labeled data scarcity problem. The State-of-the-art method uses deep neural networks, specifically a class of Recurrent Neural Network (RNN) which are Long-Short-Term-Memory networks (LSTMs) \cite{hochreiter1997long}. Deep neural networks, due to their large number of free parameters relies heavily on large annotated corpora for learning the end task. But in the real-world, it is hard to get large labeled data, mainly due to the heavy cost associated with the manual annotation. To this end, we propose a novel semi-supervised learning based RNN model, which can leverage unlabeled data  also present in abundance on social media. Through experiments we demonstrate the effectiveness of our method, achieving state-of-the-art performance in ADR mention extraction.
\end{abstract}

%
%



\maketitle

\section{Introduction}
Adverse-Drug-Reactions (ADRs) are a leading cause of mortality and morbidity in health care. In a study, it was observed that from a death count in the range of (44,000-98,000) due to medical errors, ~7000 deaths occurred due to ADRs.\footnote{http://bit.ly/2vaWF6e}. Postmarket drug surveillance is therefore required to identify such potential adverse reactions. The formal systems for postmarket surveillance can be slow and under-efficient. Studies show that ~94\% ADRs are under-reported \cite{hazell2005under}. 

Social media presents a useful platform to conduct such postmarket surveillance, given the large audience and vast reach of such platforms. Such platforms have been used for real-time information retrieval and trends tracking, including digital disease surveillance system \cite{lee2015mining}. Recent study shows that twitter has 3 times more ADRs reported than were reported through FDA. Out of ~61,000 tweets collected, 4400 had mention of ADRs as compared to 1400 ADRs reported through FDA during the same time-period \cite{Freifeld2014}. This makes Twitter a great source for building a real-time post-marketing drug safety surveillance system. However, information extraction from social media comes with its own set of challenges. Some of them are: \textbf{1)} Short nature of the text (twitter has a 142 character limit), making the language more ambiguous. \textbf{2)} Sparsity of drug-related tweets \textbf{3)} Highly colloquial language as compared to more technical and formal medical reports. 

Consider for example the tweets, '\textit{Cymbalta, you're driving me insane'; '@$<$USER$>$ Ugh, sorry. This effexor is not making me feel so awesome}'. In the first tweet, '\textit{driving me insane}' and in the second one, '\textit{not making me feel so awesome}' are ADR mentions which indicate some level of discomfort in the user's body. These tweets clearly show how information extraction from social media suffers from above-mentioned problems.

Recent work in deep learning has demonstrated its superiority over traditional hand-crafted feature based machine learning models \cite{kim2014convolutional,lee2016sequential}. However, due to a large number of free parameters, deep learning models rely heavily on large annotated dataset. In the real-world, it is often the case that labeled data is sparse, making it challenging to train such models. Semi-supervised learning based methods provide a viable solution to this. These methods rely on a small labeled data and a large unlabeled data for training. 

In this work, we present a novel semi-supervised Recurrent Neural Network (RNN) \cite{graves2012sequence} based method for ADR mention extraction, specifically leveraging a relatively large unlabeled data. We demonstrate the effectiveness of our method through experimentation on ADR mention annotated tweet corpus \cite{cocos2017deep}. Our method achieves superior results than the current state-of-the-art in ADR extraction from twitter. Our main contributions are :
\begin{itemize}
\item We propose a novel semi-supervised sequence labeling method based on RNN, specifically Long-Short-Term-Memory Network \cite{hochreiter1997long} which are known to capture long-term dependencies better than vanilla RNN. 
\item For the unsupervised learning part, we explore a novel problem of drug name prediction given context from tweets. The goal is to predict the drug-name which is masked, given it's context in the tweet. 
\item For supervised learning, we explore different word embedding initializing schemes and present results for the same.
\item We demonstrate that by training a semi-supervised model, ADR extraction performance can be improved substantially as compared to current methods. 
\item On the twitter dataset with ADR mentions annotated \cite{cocos2017deep}, our method achieves an F-score of 0.751 surpassing the current state-of-the-art method by 3.01\%. 
\end{itemize}


\section{Related Work}
The problem of ADR mention extraction falls under the category of sequence labeling problem. State-of-the-art method for sequence labeling problem is Conditional Random Fields (CRFs) \cite{lafferty2001conditional}. ADRMine \cite{nikfarjam2015pharmacovigilance}, is a CRF-based model for ADR extraction task. It uses a variety of hand-crafted features, including word context, ADR lexicon, POS-tag and word embedding based features as input to CRF. The word embedding based features are trained on a large domain-specific tweet corpus. 
The problem with the above-mentioned approach is its dependency on hand-crafted features, which is time and effort consuming. A Long-Short-Term-Memory (LSTM) network based model is proposed \cite{cocos2017deep} to get around this problem. Instead of using human-engineered features, word embedding based features are passed to a Bi-directional LSTM model which is trained to generate  a sequence of labels, given the input word sequence. State-of-the-art results are achieved, surpassing CRF-based ADRMine results. 

Some recent work also focuses on the problem of Adverse-Drug-Event (ADE) detection \cite{lee2017adverse,huynh2016adverse}. The goal is to identify whether there is an Adverse-Drug-Event mentioned in the tweet based on its textual content.  

\section{ADR-mention extraction using Semi-supervised Bi-directional LSTM}
In this section, we present our approach for ADR extraction. Our method is based on a semi-supervised learning method which operates in two phases: \textbf{1) Unsupervised learning:} In this phase, we train a Bidirectional LSTM model to predict the drug name given its context in the tweet. As training data for this task, we select tweets with exactly one mention of any prescription drug. Since we already know the drug name beforehand, it doesn't need any annotation effort. \textbf{2) Supervised learning:} In this phase, we use the same bidirectional LSTM model from phase 1 and (re)train it to predict the sequence of labels, given the tweet text.

\subsection{Unsupervised learning}
For this phase, we choose a novel task of drug name prediction from its context in the tweet. For training data, we use a large collection of tweets with exactly one mention of the drug name in them. Since we are predicting the drug name from a tweet which is already present in it, in order to avoid the network to learn a trivial function which maps drug-name in input to drug-name in output without considering the context in account, we mask the drug-name in the tweet with a dummy token. For feature-extraction, we use a Bidirectional LSTM model. The model takes as input, a sequence of continuous word vectors as input and predicts a corresponding sequence of word vectors as output. The equations governing the dynamics of LSTMs are defined as follows:

\begin{equation}
	\begin{split}
		&  \vec g^u = \sigma( W^u * \vec h_{t-1} +  I^u * \vec \x_t) \\
		& \vec g^f = \sigma( W^f * \vec h_{t-1} +  I^f * \vec \x_t) \\
        &  \vec g^c = \tanh( W^c * \vec h_{t-1} +  I^c * \vec \x_t) \\
		&  \vec m_t =  \vec g^f \odot  +  \vec g^u \odot \vec  g^c \\
		&  \vec g^o = \sigma(W^o *  \vec h_{t-1} +  I^o * \vec \x_t) \\
		&  \vec h_t = \tanh( \vec g^o \odot \vec m_{t-1}) 
	\end{split}
\end{equation}
here $\sigma$ is the logistic sigmoid function, $\mathbf W^u, \mathbf W^f, \mathbf W^o, \mathbf W^c$ are recurrent weight matrices and $\mathbf I^u, \mathbf I^f, \mathbf I^o, \mathbf I^c$ are projection matrices. In a conventional LSTM, the sequence order is from left to right. In Bidirectional LSTM, two sequence directions are considered, one from left to right and the other one opposite to it. The final hidden layers activation is the concatenation of vectors from both directions. Mathematically, 
\begin{equation}
   \mathbf h_t = [\vec h_t ; \cev h_t]
\end{equation}
To generate the final representation of the tweet, average-pooling is applied over all hidden state vectors. 
\begin{equation}
   \mathbf h = \sum_{t=1}^{T} \mathbf h_t
\end{equation}
where T is the maximum time-step. Finally a softmax transformation is applied to generate a probability distribution over all drug-names followed by categorical cross-entropy loss.

\begin{table*}[t]
\centering
\label{Main}
 \begin{tabular}{||c c c c||}
 \hline
 Method & F1-Score & Precision & Recall \\ [1ex] 
 \hline\hline
 \textbf{Baseline} \cite{cocos2017deep} & 0.729 $\pm$ 0.027 & 0.695 $\pm$ 0.109  & \textbf{0.776 $\pm$ 0.121} \\
 \hline
 Baseline (with adam optimizer) & 0.737 $\pm$ 0.308  & 0.707 $\pm$ 0.096 & 0.774 $\pm$ 0.08 \\ 
 Semi-Supervised ADR extraction & \textbf{0.751 $\pm$ 0.036} & \textbf{0.731 $\pm$ 0.035} & 0.774 $\pm$ 0.073 \\ 
 \hline
 \end{tabular}
 \caption{Performance of various deep neural network methods on ADR extraction task. Results are averaged over 10 trails, and are presented with std. deviation}
\end{table*}
\subsection{Supervised Sequence Classification}
For this phase, we reuse the Bidirectional LSTM trained from the previous phase following the setup similar to state-of-the-art \cite{cocos2017deep}. At each time-step of the sequence, a softmax layer is applied which predicts a probability distribution over sequence labels. Formally, 

\begin{equation}
\mathbf y_t = softmax(\mathbf W \mathbf h +  \mathbf b) \\
\end{equation}
here $\mathbf W$ and $\mathbf b$ are weight matrices for the softmax layer. The final loss for the sequence labeling is sum of categorical cross-entropy loss at each time-step. The hidden state $\mathbf h$ and the parameters $\mathbf W^u$, $\mathbf W^f$, $\mathbf W^o$, $\mathbf W^c$, $\mathbf I^u$, $\mathbf I^f$, $\mathbf I^o$, $\mathbf I^c$ are shared during training both phases. 

The intuition around the unsupervised task is that the network can learn the textual context where drug names appear, which can help in identifying Adverse Drug Reactions from drugs. 
\section{Experiments}

\subsection{Dataset}
We use the twitter dataset annotated with ADR mention collected during the period of 2007-2010. Tweets were collected using 81 drug names as keyword search terms. In the original dataset, a total of ~960 tweets are annotated with ADR mentions. Due to Twitter's search APIs license, only tweet ids were released. Out of the total of ~960, we collected a total of ~645 tweets using Python library tweepy\footnote{https://github.com/tweepy/tweepy}. According to the given train-test split, 470 tweets are used for training and 170 tweets for testing.  

For the unlabeled dataset, we used the Twitter's search API \footnote{https://dev.twitter.com/rest/public/search} with the drug names used in the original study as keyword search terms \footnote{http://diego.asu.edu/Publications/ADRMine.html \newline Some example drug names used as keywords are: \textbf{humira, dronedarone, lamictal, 
pradaxa, paxil, zoledronic acid, trazodone, enbrel, cymbalta, quetiapine  }}. We crawled the tweets over a period of two months. For the sake of simplicity, we removed the tweets with more than one drug mentions, , resulting in a total of 0.1 Million tweets.

\subsection{Implementation Details}
We use Keras\footnote{https://keras.io/} for implementation. For text pre-processing, we applied several pre-processing steps, which are :
\begin{itemize}
\item \texttt{Normalizing HTML links and user-mentions:}We replaced all HTML link mentions with the token "$<$LINK$>$". Similarly, we replaced all user handle mentions (for ex. @JonDoe) with the token "$<$USER$>$".
\item \texttt{Special Character Removal:} We removed all punctuations and special symbols like '\#' from tweets.
\item \texttt{Emoticons Removal:} We removed all emoticons, in general all non-ascii characters which are special types of emoticons.
\item \texttt{Stop-word and rare words removal:} We removed all stop-words and set the vocabulary size to top-15000 most frequent words in the corpus.
\end{itemize}
We used the word2vec \cite{mikolov2013distributed} embeddings trained on a large generic twitter corpus \cite{godin2015multimedia} as input to the model. Word vector dimension is set to 400. BiLSTM parameters are set to the best reported setting from \cite{cocos2017deep}, with hidden unit's dimension equal to 500. For training the supervised model, we use the adam optimizer \cite{kingma2014adam} with batch-size equal to 1 and for training the unsupervised model, we used the batch adam optimizer \cite{kingma2014adam} with batch-size set to 128. The supervised model was trained for a total of 5 epochs, and unsupervised model trained for 30 epochs.

\subsection{Results}
To convert ADR extraction problem into sequence labeling problem, we need to assign annotated entities with appropriate tag representations. We follow IO encoding scheme where each word belongs to either of the following categories: \textbf{(1)} I-ADR (inside ADR) \textbf{(2)} I-Indication (inside Indication ) \textbf{(3)} O (Outside any mention) \textbf{(4)} $<$PAD$>$ (if the word is padding token). It should be noted that, similar to the baseline method \cite{cocos2017deep} we report the performance on the ADR label only. This is because the number of Indication annotations are very less in number\footnote{45 in training, 16 in testing}. An example tweet annotated with IO-encoding:\textit{@BLENDOS\textsubscript{O} Lamictal\textsubscript{O} and\textsubscript{O} trileptal\textsubscript{O} and\textsubscript{O} seroquel\textsubscript{O} of\textsubscript{O} course\textsubscript{O} the\textsubscript{O} seroquel\textsubscript{O} I\textsubscript{O} take\textsubscript{O} in\textsubscript{O} severe\textsubscript{O} situations\textsubscript{O} because\textsubscript{O} weight\textsubscript{I-ADR} gain\textsubscript{I-ADR} is\textsubscript{O} not\textsubscript{O}  cool\textsubscript{O}} For performance evaluation we use approximate-matching \cite{tsai2006various}, which is used popularly in biomedical entity extraction tasks \cite{cocos2017deep,nikfarjam2015pharmacovigilance}. We report the F1-score, Precision and Recall computed using approximate matching as follows:
\begin{equation}
\text{Precision} = \frac{\text{\#ADR approximately matched}}{\text{\#ADR spans predicted}}
\end{equation}
\begin{equation}
\text{Recall}= \frac{\text{\#ADR approximately matched}}{\text{\#ADR 
spans in total}}
\end{equation}
Table 1 presents the results of our approach along with comparisons. Since the number of tweets used for training and testing differs from the one used in baseline \cite{cocos2017deep}, we re-ran their model using the source-code released by them\footnote{https://github.com/chop-dbhi/twitter-adr-blstm}. It should be noted that the original model used RMSProp \cite{tieleman2012lecture} as an optimizer, so for a fair comparison we also report the baseline results with optimizer as adam instead of RMSProp. Replacing RMSProp with adam, although gives an improvement over the original baseline, still under-performs our method . Our approach gives the state-of-the-art results, giving an improvement of 2.97\%F1 over the original baseline and an improvement of 1.88\% F1 over the re-implemented baseline. 

\subsection{Analysis}
\begin{table*}[ht!]
\centering
\label{Main}
 \begin{tabular}{||c c c c||} 
 \hline
 Method & F1-Score & Precision & Recall \\ [1ex] 
 \hline\hline
 SS-BLSTM (with drug mask removed) & \textbf{0.747 $\pm$ 0.035} & 0.723 $\pm$ 0.104 & \textbf{0.780 $\pm$ 0.105} \\
 SS-BLSTM (with labeled tweets dictionary only) & 0.745 $\pm$ 0.034  & \textbf{0.727 $\pm$ 0.075} & 0.769 $\pm$ 0.093 \\ 
 SS-BLSTM (with GoogleNews\footnote{https://code.google.com/archive/p/word2vec/} vectors) & 0.736 $\pm$ 0.036 & 0.708 $\pm$ 0.091 & 0.774 $\pm$ 0.115 \\
 SS-BLSTM (with medical embeddings) & 0.673 $\pm$ 0.025 & 0.642 $\pm$ 0.085 & 0.716 $\pm$ 0.116 \\
\hline
 \end{tabular}
 \caption{Performance comparison of Semi-Supervised BiLSTM (SS-BLSTM) under different word embedding initialization settings and different unlabeled data settings. Results are reported with average over 10 trails along with the std. deviation}
\end{table*}

\subsubsection{Effect of drug-mask}
For the unsupervised learning phase, we select the task of drug-name prediction given its context. In order to avoid the network learning a degenerate function which maps input drug-name to output drug-name, we mask all drug-names in input with a single token. In order to verify this, we report the accuracy results without the drug-mask, i.e. with drug-name included in the input. The result is presented in Table 2. It is clear that removing the drug mask from input degrades the end-performance by 0.535\% in F-score. This further validates our claim that masking the drug-names is effective.

\subsubsection{Effect of embeddings and dictionary}
We experiment with word embeddings trained on different corpus to observe its effect on the end-performance. We experiment with embeddings trained on part of Google News Dataset, which consists of around 100 billion words\footnote{https://code.google.com/archive/p/word2vec/}. It can be observed that using Google News Corpus trained embeddings degrade the performance by 2.038\% in F-score. This is due to the fact that these embeddings are trained on a large News Corpus, which is more grammatically sound and formal than the raw social media posts. Conceptually, the shift in the lexical data distribution of the News corpus as compared to tweets containing ADR causes the degradation in performance. We also experiment with word embeddings trained on a large medical-concept terms related tweet corpus\footnote{https://zenodo.org/record/27354\#.WWYph1ekW4A} \cite{limsopatham2015adapting}. Intuitively, embeddings trained on similar domain (medical in this case) should perform better, but  surprisingly it performs worst amongst all methods. The generic embeddings trained on large tweet corpus captures potentially large variation of semantics and linguistic properties of text and due to the free-style nature of writing on social media, this helps more than domain-knowledge, as captured by medical-domain trained embeddings.

We also experimented with a different vocabulary initialization. In our proposed formulation, we construct vocabulary from both unlabeled and labeled corpus, resulting in a larger vocabulary size. When experimented with a restricted vocabulary (only from labeled training data), we observe that the F1-score drops by 0.8\%. This suggests the use of a larger vocabulary with more coverage in similar settings. 

\section{Conclusions}
We present a novel semi-supervised Bi-directional LSTM based model for ADR mention extraction. We evaluate our method on an annotated twitter corpus. By leveraging potentially large unlabeled corpus, our method outperforms the state-of-the-art method by 3.01\% in F1-score. 

We also demonstrate that word embeddings trained on a large domain-agnostic twitter corpus performs better than more popular Google News Corpus trained word-embeddings and surprisingly even better than medical domain-specific word embeddings trained on tweets, which suggests that  language structure and semantics is more important in downstream information extraction tasks, compared to domain knowledge. 

In future, we will explore drug and side-effect (adverse-effect) mention relation along with ADR extraction and explore if both can be formulated in a multi-task learning setup.

\bibliographystyle{ACM-Reference-Format}
\bibliography{sample-bibliography} 


\begin{thebibliography}{00}


\ifx \showCODEN    \undefined \def \showCODEN     #1{\unskip}     \fi
\ifx \showDOI      \undefined \def \showDOI       #1{#1}\fi
\ifx \showISBNx    \undefined \def \showISBNx     #1{\unskip}     \fi
\ifx \showISBNxiii \undefined \def \showISBNxiii  #1{\unskip}     \fi
\ifx \showISSN     \undefined \def \showISSN      #1{\unskip}     \fi
\ifx \showLCCN     \undefined \def \showLCCN      #1{\unskip}     \fi
\ifx \shownote     \undefined \def \shownote      #1{#1}          \fi
\ifx \showarticletitle \undefined \def \showarticletitle #1{#1}   \fi
\ifx \showURL      \undefined \def \showURL       {\relax}        \fi
\providecommand\bibfield[2]{#2}
\providecommand\bibinfo[2]{#2}
\providecommand\natexlab[1]{#1}
\providecommand\showeprint[2][]{arXiv:#2}

\bibitem[\protect\citeauthoryear{Cocos, Fiks, and Masino}{Cocos
  et~al\mbox{.}}{2017}]%
        {cocos2017deep}
\bibfield{author}{\bibinfo{person}{Anne Cocos}, \bibinfo{person}{Alexander~G
  Fiks}, {and} \bibinfo{person}{Aaron~J Masino}.}
  \bibinfo{year}{2017}\natexlab{}.
\newblock \showarticletitle{Deep learning for pharmacovigilance: recurrent
  neural network architectures for labeling adverse drug reactions in Twitter
  posts}.
\newblock \bibinfo{journal}{{\em Journal of the American Medical Informatics
  Association\/}} (\bibinfo{year}{2017}), \bibinfo{pages}{ocw180}.
\newblock


\bibitem[\protect\citeauthoryear{Freifeld, Brownstein, Menone, Bao, Filice,
  Kass-Hout, and Dasgupta}{Freifeld et~al\mbox{.}}{2014}]%
        {Freifeld2014}
\bibfield{author}{\bibinfo{person}{Clark~C. Freifeld}, \bibinfo{person}{John~S.
  Brownstein}, \bibinfo{person}{Christopher~M. Menone}, \bibinfo{person}{Wenjie
  Bao}, \bibinfo{person}{Ross Filice}, \bibinfo{person}{Taha Kass-Hout}, {and}
  \bibinfo{person}{Nabarun Dasgupta}.} \bibinfo{year}{2014}\natexlab{}.
\newblock \showarticletitle{Digital Drug Safety Surveillance: Monitoring
  Pharmaceutical Products in Twitter}.
\newblock \bibinfo{journal}{{\em Drug Safety\/}} \bibinfo{volume}{37},
  \bibinfo{number}{5} (\bibinfo{date}{01 May} \bibinfo{year}{2014}),
  \bibinfo{pages}{343--350}.
\newblock
\showISSN{1179-1942}
\showDOI{%
\url{https://doi.org/10.1007/s40264-014-0155-x}}


\bibitem[\protect\citeauthoryear{Godin, Vandersmissen, De~Neve, and Van~de
  Walle}{Godin et~al\mbox{.}}{2015}]%
        {godin2015multimedia}
\bibfield{author}{\bibinfo{person}{Fr{\'e}deric Godin},
  \bibinfo{person}{Baptist Vandersmissen}, \bibinfo{person}{Wesley De~Neve},
  {and} \bibinfo{person}{Rik Van~de Walle}.} \bibinfo{year}{2015}\natexlab{}.
\newblock \showarticletitle{Multimedia lab@ acl w-nut ner shared task: named
  entity recognition for twitter microposts using distributed word
  representations}.
\newblock \bibinfo{journal}{{\em ACL-IJCNLP\/}}  \bibinfo{volume}{2015}
  (\bibinfo{year}{2015}), \bibinfo{pages}{146--153}.
\newblock


\bibitem[\protect\citeauthoryear{Graves}{Graves}{2012}]%
        {graves2012sequence}
\bibfield{author}{\bibinfo{person}{Alex Graves}.}
  \bibinfo{year}{2012}\natexlab{}.
\newblock \showarticletitle{Sequence transduction with recurrent neural
  networks}.
\newblock \bibinfo{journal}{{\em arXiv preprint arXiv:1211.3711\/}}
  (\bibinfo{year}{2012}).
\newblock


\bibitem[\protect\citeauthoryear{Hazell and Shakir}{Hazell and Shakir}{2005}]%
        {hazell2005under}
\bibfield{author}{\bibinfo{person}{Lorna Hazell} {and} \bibinfo{person}{Saad~Aw
  Shakir}.} \bibinfo{year}{2005}\natexlab{}.
\newblock \showarticletitle{Under-reporting of Adverse Drug Reactions: A
  Systematic Review}.
\newblock \bibinfo{journal}{{\em Pharmacoepidemiology and Drug Safety\/}}
  \bibinfo{volume}{14} (\bibinfo{year}{2005}), \bibinfo{pages}{S184--S185}.
\newblock


\bibitem[\protect\citeauthoryear{Hochreiter and Schmidhuber}{Hochreiter and
  Schmidhuber}{1997}]%
        {hochreiter1997long}
\bibfield{author}{\bibinfo{person}{Sepp Hochreiter} {and}
  \bibinfo{person}{J{\"u}rgen Schmidhuber}.} \bibinfo{year}{1997}\natexlab{}.
\newblock \showarticletitle{Long short-term memory}.
\newblock \bibinfo{journal}{{\em Neural computation\/}} \bibinfo{volume}{9},
  \bibinfo{number}{8} (\bibinfo{year}{1997}), \bibinfo{pages}{1735--1780}.
\newblock


\bibitem[\protect\citeauthoryear{Huynh, He, Willis, and R{\"u}ger}{Huynh
  et~al\mbox{.}}{2016}]%
        {huynh2016adverse}
\bibfield{author}{\bibinfo{person}{Trung Huynh}, \bibinfo{person}{Yulan He},
  \bibinfo{person}{Allistair Willis}, {and} \bibinfo{person}{Stefan
  R{\"u}ger}.} \bibinfo{year}{2016}\natexlab{}.
\newblock \showarticletitle{Adverse drug reaction classification with deep
  neural networks}.
\newblock  (\bibinfo{year}{2016}).
\newblock


\bibitem[\protect\citeauthoryear{Kim}{Kim}{2014}]%
        {kim2014convolutional}
\bibfield{author}{\bibinfo{person}{Yoon Kim}.} \bibinfo{year}{2014}\natexlab{}.
\newblock \showarticletitle{Convolutional neural networks for sentence
  classification}.
\newblock \bibinfo{journal}{{\em arXiv preprint arXiv:1408.5882\/}}
  (\bibinfo{year}{2014}).
\newblock


\bibitem[\protect\citeauthoryear{Kingma and Ba}{Kingma and Ba}{2014}]%
        {kingma2014adam}
\bibfield{author}{\bibinfo{person}{Diederik Kingma} {and}
  \bibinfo{person}{Jimmy Ba}.} \bibinfo{year}{2014}\natexlab{}.
\newblock \showarticletitle{Adam: A method for stochastic optimization}.
\newblock \bibinfo{journal}{{\em arXiv preprint arXiv:1412.6980\/}}
  (\bibinfo{year}{2014}).
\newblock


\bibitem[\protect\citeauthoryear{Lafferty, McCallum, Pereira,
  et~al\mbox{.}}{Lafferty et~al\mbox{.}}{2001}]%
        {lafferty2001conditional}
\bibfield{author}{\bibinfo{person}{John Lafferty}, \bibinfo{person}{Andrew
  McCallum}, \bibinfo{person}{Fernando Pereira}, {et~al\mbox{.}}}
  \bibinfo{year}{2001}\natexlab{}.
\newblock \showarticletitle{Conditional random fields: Probabilistic models for
  segmenting and labeling sequence data}.
\newblock  (\bibinfo{year}{2001}).
\newblock


\bibitem[\protect\citeauthoryear{Lee and Dernoncourt}{Lee and
  Dernoncourt}{2016}]%
        {lee2016sequential}
\bibfield{author}{\bibinfo{person}{Ji~Young Lee} {and} \bibinfo{person}{Franck
  Dernoncourt}.} \bibinfo{year}{2016}\natexlab{}.
\newblock \showarticletitle{Sequential Short-Text Classification with Recurrent
  and Convolutional Neural Networks}. In \bibinfo{booktitle}{{\em Proceedings
  of NAACL-HLT}}. \bibinfo{pages}{515--520}.
\newblock


\bibitem[\protect\citeauthoryear{Lee, Agrawal, and Choudhary}{Lee
  et~al\mbox{.}}{2015}]%
        {lee2015mining}
\bibfield{author}{\bibinfo{person}{Kathy Lee}, \bibinfo{person}{Ankit Agrawal},
  {and} \bibinfo{person}{Alok Choudhary}.} \bibinfo{year}{2015}\natexlab{}.
\newblock \showarticletitle{Mining social media streams to improve public
  health allergy surveillance}. In \bibinfo{booktitle}{{\em ASONAM}}.
  \bibinfo{pages}{815--822}.
\newblock


\bibitem[\protect\citeauthoryear{Lee, Qadir, Hasan, Datla, Prakash, Liu, and
  Farri}{Lee et~al\mbox{.}}{2017}]%
        {lee2017adverse}
\bibfield{author}{\bibinfo{person}{Kathy Lee}, \bibinfo{person}{Ashequl Qadir},
  \bibinfo{person}{Sadid~A Hasan}, \bibinfo{person}{Vivek Datla},
  \bibinfo{person}{Aaditya Prakash}, \bibinfo{person}{Joey Liu}, {and}
  \bibinfo{person}{Oladimeji Farri}.} \bibinfo{year}{2017}\natexlab{}.
\newblock \showarticletitle{Adverse Drug Event Detection in Tweets with
  Semi-Supervised Convolutional Neural Networks}. In \bibinfo{booktitle}{{\em
  WWW}}. \bibinfo{pages}{705--714}.
\newblock


\bibitem[\protect\citeauthoryear{Limsopatham and Collier}{Limsopatham and
  Collier}{2015}]%
        {limsopatham2015adapting}
\bibfield{author}{\bibinfo{person}{Nut Limsopatham} {and}
  \bibinfo{person}{Nigel Collier}.} \bibinfo{year}{2015}\natexlab{}.
\newblock \showarticletitle{Adapting phrase-based machine translation to
  normalise medical terms in social media messages}.
\newblock \bibinfo{journal}{{\em arXiv preprint arXiv:1508.02285\/}}
  (\bibinfo{year}{2015}).
\newblock


\bibitem[\protect\citeauthoryear{Mikolov, Sutskever, Chen, Corrado, and
  Dean}{Mikolov et~al\mbox{.}}{2013}]%
        {mikolov2013distributed}
\bibfield{author}{\bibinfo{person}{Tomas Mikolov}, \bibinfo{person}{Ilya
  Sutskever}, \bibinfo{person}{Kai Chen}, \bibinfo{person}{Greg~S Corrado},
  {and} \bibinfo{person}{Jeff Dean}.} \bibinfo{year}{2013}\natexlab{}.
\newblock \showarticletitle{Distributed representations of words and phrases
  and their compositionality}. In \bibinfo{booktitle}{{\em NIPS}}.
  \bibinfo{pages}{3111--3119}.
\newblock


\bibitem[\protect\citeauthoryear{Nikfarjam, Sarker, O’Connor, Ginn, and
  Gonzalez}{Nikfarjam et~al\mbox{.}}{2015}]%
        {nikfarjam2015pharmacovigilance}
\bibfield{author}{\bibinfo{person}{Azadeh Nikfarjam}, \bibinfo{person}{Abeed
  Sarker}, \bibinfo{person}{Karen O’Connor}, \bibinfo{person}{Rachel Ginn},
  {and} \bibinfo{person}{Graciela Gonzalez}.} \bibinfo{year}{2015}\natexlab{}.
\newblock \showarticletitle{Pharmacovigilance from social media: mining adverse
  drug reaction mentions using sequence labeling with word embedding cluster
  features}.
\newblock \bibinfo{journal}{{\em Journal of the American Medical Informatics
  Association\/}} \bibinfo{volume}{22}, \bibinfo{number}{3}
  (\bibinfo{year}{2015}), \bibinfo{pages}{671--681}.
\newblock


\bibitem[\protect\citeauthoryear{Tieleman and Hinton}{Tieleman and
  Hinton}{2012}]%
        {tieleman2012lecture}
\bibfield{author}{\bibinfo{person}{Tijmen Tieleman} {and}
  \bibinfo{person}{Geoffrey Hinton}.} \bibinfo{year}{2012}\natexlab{}.
\newblock \showarticletitle{Lecture 6.5-rmsprop: Divide the gradient by a
  running average of its recent magnitude}.
\newblock \bibinfo{journal}{{\em COURSERA: Neural networks for machine
  learning\/}} \bibinfo{volume}{4}, \bibinfo{number}{2} (\bibinfo{year}{2012}),
  \bibinfo{pages}{26--31}.
\newblock


\bibitem[\protect\citeauthoryear{Tsai, Wu, Chou, Lin, He, Hsiang, Sung, and
  Hsu}{Tsai et~al\mbox{.}}{2006}]%
        {tsai2006various}
\bibfield{author}{\bibinfo{person}{Richard Tzong-Han Tsai},
  \bibinfo{person}{Shih-Hung Wu}, \bibinfo{person}{Wen-Chi Chou},
  \bibinfo{person}{Yu-Chun Lin}, \bibinfo{person}{Ding He},
  \bibinfo{person}{Jieh Hsiang}, \bibinfo{person}{Ting-Yi Sung}, {and}
  \bibinfo{person}{Wen-Lian Hsu}.} \bibinfo{year}{2006}\natexlab{}.
\newblock \showarticletitle{Various criteria in the evaluation of biomedical
  named entity recognition}.
\newblock \bibinfo{journal}{{\em BMC bioinformatics\/}} \bibinfo{volume}{7},
  \bibinfo{number}{1} (\bibinfo{year}{2006}), \bibinfo{pages}{92}.
\newblock


\end{thebibliography}

\end{document}